\documentclass{mem}
\usepackage{natbib}\usepackage{txfonts}\usepackage{balance}
\usepackage{graphicx}
\usepackage[a4paper,breaklinks,dvipdfm]{hyperref}
\idline{75}{282}
\begin{document}
\def\teff{$T\rm_{eff }$}
\def\kms{$\mathrm {km s}^{-1}$}

\title{Dynamics of supernova remnants\\
in the Galactic Centre}

   \subtitle{}

\author{
Elisa \,Bortolas\inst{1,2} 
\and Michela \, Mapelli\inst{1}
\and Mario \, Spera\inst{1}
          }

\institute{
INAF, Osservatorio Astronomico di Padova, Vicolo dell'Osservatorio 5, I-35122, Padova, Italy
\and
Dipartimento di Fisica e Astronomia ``Galileo Galilei'', Universit\`a di Padova, Vicolo dell'Osservatorio 3, 35122 Padova, Italy\\
\email{elisa.bortolas@oapd.inaf.it}
}

\authorrunning{Bortolas }

\titlerunning{Dynamics of Supernova remnants in the Galactic Centre}

\abstract{
The Galactic centre (GC) is a unique place to study the extreme dynamical processes occurring near a super-massive black hole (SMBH). Here we simulate a large set of binaries orbiting the SMBH while the primary member undergoes a supernova (SN) explosion, in order to study the impact of SN kicks on the orbits of stars and dark remnants in the GC. We find that SN explosions are efficient in scattering neutron stars  and other light stars on new (mostly eccentric) orbits, while black holes (BHs) tend to retain memory of the orbit of their progenitor star. SN kicks are thus unable to eject BHs from the GC: a cusp of dark remnants may be lurking in the central parsec of our Galaxy.
}
\maketitle{}

\section{Introduction}
The Galactic centre (GC) is an ideal laboratory to study the extreme dynamical processes occurring in  proximity of a super-massive black hole (SMBH). The GC, which is the  sole galactic nucleus where we can genuinely resolve single stars, appears to be a very crowded and puzzling place (see \citealt{mapelli2016} for a recent review): 
 (i) a population of $\sim$ 30 young B stars  (the S-stars) lies within  0.04 pc from the SMBH, where star formation should be inhibited;
(ii) the faint dusty objects G1 and G2 have recently been spotted in the vicinity of the SMBH, but their nature is still unclear; (iii) late-type red giant stars,  expected to trace the old relaxed stellar population, show a flattened density profile, in contradiction with the cuspy distribution expected in proximity of a SMBH.

Here, we investigate the possibility that supernova (SN) kicks occurring in binary systems affect the orbits of stars and stellar remnants in the GC, by means of direct $N$-body simulations. 

\section{Methods}
We perform a suite of 10k simulations of  three-body systems, composed of the SMBH and the stellar binary. Our work is motivated by the fact that most massive stars are found in binary systems. 
We set the initial distribution of mass and orbital parameters of the simulated binaries in agreement with recent observations of the clockwise (CW) disc \citep[e.g.][]{yelda2014}.  We start our simulations when the primary member of the binary undergoes an SN explosion. The prescriptions for SN kicks are the same as in  \citet{spera2015}. 
Each three-body  system is evolved for 1 Myr by means of a regularized code \citep[Mikkola algorithmic regularization,][]{mikkola93} in the environment of \texttt{HiGPUs} \citep[]{higpus}. More details about our simulations will be provided in Bortolas et al., in preparation.

\section{Results and conclusions}

\begin{figure}[]
\resizebox{\hsize}{!}{\includegraphics[clip=true,trim={0cm -.2cm 7.5cm 16cm},width=.7\columnwidth]{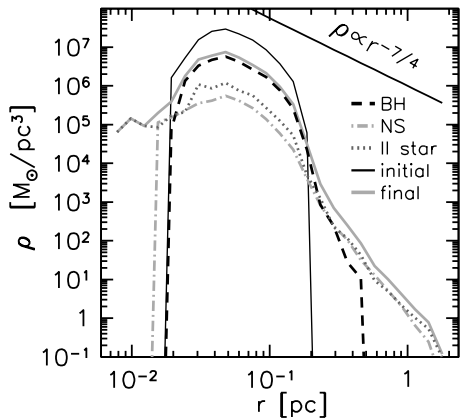} }
\caption{
\footnotesize
Density distribution of simulated stars and remnants. Solid black (grey) line: final (initial) distribution of all simulated objects. Dashed line: final distribution of BHs. Dot-dashed line: final distribution of NSs. Dotted line: final distribution of the secondary stars. BHs are the most important population inside $0.2$ pc, while NSs and secondary stars end up at larger distances (up to 2 pc). The top-right line represents the density profile expected for a relaxed cusp around a SMBH.
}
\label{dens}
\end{figure}
Figure  \ref{dens} shows the density profile of the simulated binaries after 1 Myr, distinguishing between black holes (BHs), neutron stars (NSs) and stellar companions. 
BHs are the main component in the range $\sim 0.01-0.2$ pc: their radial distribution is similar to the initial one, implying that most BHs are not ejected by SN kicks and end-up forming a hidden dark cusp. 
NSs and secondary stars, being lighter, are more easily scattered by the SN explosion and may spread in the innermost and outermost regions of the GC. We propose that the dearth of NSs in the GC, suggested by observations, might be due to SN kicks.

Most (99\%) binaries where the SN remnant is a NS break because of the kick, whereas  57\% of BHs remain bound to their stellar companion. 
We  estimate that $\lesssim$35\% of the BH binaries are expected to undergo mass transfer via Roche-lobe overflow. 

\begin{figure}[]
\resizebox{\hsize}{!}{\includegraphics[clip=true,trim={.6cm .5cm 8cm 0.4cm},width=\columnwidth]{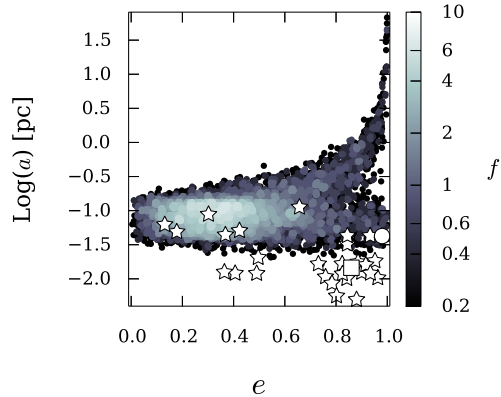} }
\caption{
\footnotesize
Eccentricity ($e$) versus the semi-major axis ($a$) of the kicked secondary stars. The symbols mark the values of $e$ and $a$ for G1 (square), G2 (open circle), and for the S-cluster members (stars). The color bar represents the points density.}
\label{scatt}
\end{figure}

Figure~\ref{scatt} shows the semi-major axis and eccentricity of the simulated secondary stars, compared to those of the S-stars and the G1 and G2 objects. We note that the orbital properties of several simulated low-mass stars match those of the G2 object, suggesting a new scenario to explain the origin of G2.

\bibliographystyle{aa}

\end{document}